\documentclass[10pt]{article} 
\usepackage{amsfonts,amssymb,slashed,makeidx,latexsym,setspace}
\usepackage{graphicx,graphics,amsmath,amssymb,epsf,rotate,subfigure} 
\usepackage{times,cite,color}
\begin{document}
\begin{flushright}
SINP/TNP/2010/12,~~TH-DO 10/10
\end{flushright}

\vskip 10pt

\begin{center}
  {\Large \bf Exotic Higgs boson decay modes as a harbinger of $S_3$ 
flavor symmetry} \\
  \vspace*{1cm} \renewcommand{\thefootnote}{\fnsymbol{footnote}} { {\sf Gautam
      Bhattacharyya ${}^{1)}$}, {\sf Philipp Leser ${}^{2)}$}},
  and {\sf Heinrich P\"as ${}^{2)}$} \\
  \vspace{10pt} {\small ${}^{1)}$ {\em Saha Institute of Nuclear Physics, 1/AF
      Bidhan Nagar, Kolkata 700064, India }}\\
  {\small ${}^{2)}$ {\em Fakult\"at f\"ur Physik, Technische Universit\"at
      Dortmund, 44221 Dortmund, Germany}} \normalsize
\end{center}

\begin{abstract}
  Discrete symmetries employed to explain flavor mixing and mass hierarchies
  can be associated with an enlarged scalar sector which might lead to
  exotic Higgs decay modes. In this paper, we explore such a possibility in a
  scenario with $S_3$ flavor symmetry which requires three scalar $SU(2)$
  doublets.  The spectrum is fixed by minimizing the scalar potential, and we
  observe that the symmetry of the model leads to tantalizing Higgs decay
  modes potentially observable at the CERN Large Hadron Collider (LHC).

\vskip 5pt \noindent
\texttt{PACS Nos:~14.80.Ec, 11.30.Hv} \\
\texttt{Key Words:~~Higgs boson, flavor symmetry}
\end{abstract}

\setcounter{footnote}{0}
\renewcommand{\thefootnote}{\arabic{footnote}}

\section{Introduction}
Flavor models based on discrete symmetries are often used to address
issues like large/small mixing and mild/strong mass hierarchies in the
lepton/quark sector (for reviews, see, e.g.~\cite{discrete_revs}).
The permutation group $S_3$ is an attractive such candidate which was
introduced in \cite{Pakvasa:1977in} and explored further in
\cite{s3pheno,Chen:2004rr}.  In this paper we study the exciting
prospect that such flavor models can predict {\em enlarged Higgs
  sectors with non-standard couplings to fermions and gauge bosons},
although such a nonsupersymmetric extension does not provide any
additional stability to the potential. The main motivation here arises
from flavor issues. Its supersymmetrization would indeed bring in
quantum stability, but in this work we stick to a minimal
nonsupersymmetric $S_3$ flavor scenario.

The motivation for choosing $S_3$ is that it is the smallest
non-abelian discrete symmetry group that contains a
\mbox{2-dimensional} irreducible representation which can connect two
maximally mixed generations.  It is the symmetry group of an
equilateral triangle and has three irreducible representations:
$\mathbf{1},\mathbf{1'}$ and $\mathbf{2}$, with multiplication rules:
$\mathbf{2}\times\mathbf{2}=\mathbf{1}+\mathbf{1'}+\mathbf{2}$ and
$\mathbf{1'}\times \mathbf{1'}=\mathbf{1}$.  Besides facilitating
maximal mixing through its doublet representation, $S_3$ provides two
inequivalent singlet representations which play a crucial role in
reproducing fermion masses and mixing. To accomplish the latter, three
scalar $SU(2)$ doublets are introduced, which couple to the fermions
as dictated by $S_3$ symmetry. It so happens that large mixing among
up- and down-type quarks cancel each other in the Cabibbo
matrix. Neutrino masses are separately generated by a type-II see-saw
mechanism using scalar $SU(2)$ triplets \cite{Mohapatra:1979ia}, so
that the mismatch between the large mixing of the charged leptons and
the diagonal neutrino masses translates directly into the
Pontecorvo-Maki-Nakagawa-Sakata matrix. In this paper we do not deal
with those triplets, but explore the following avenues: ($i$)
minimization of the scalar potential with three scalar $SU(2)$
doublets, two of which form an $S_3$ doublet and the third an $S_3$
singlet, ($ii$) the gauge and Yukawa interactions of the neutral
scalars, and ($iii$) different nonstandard production and decay modes
of the neutral CP-even scalars leading to the possibility of their
detection at the LHC.

For definiteness, we study the $S_3$ model pursued in
\cite{Chen:2004rr} to explain the leptonic flavor structure.  We
concentrate on the complementary aspects by exploring the scalar
sector.  The assignments of the fermion and scalar fields are as
follows:
\begin{align}
	\label{eq:multipletassignments}
	(L_\mu,L_\tau)&\in \mathbf{2} & L_e, e^c,\mu^c
&\in \mathbf{1}
        & \tau^c&\in \mathbf{1'} \, , \nonumber \\
	(Q_2,Q_3)&\in\mathbf{2} & Q_1, u^c,c^c,d^c,s^c
&\in\mathbf{1}
        & b^c,t^c&\in\mathbf{1'} \, , \\
	(\phi_1,\phi_2)&\in \mathbf{2} & \phi_3&\in\mathbf{1} \, \nonumber, 
\end{align}
where the notations are standard and self-explanatory.  The vacuum
expectation values (VEVs) of the three scalar doublets $\phi_{1,2,3}$
induce spontaneous electroweak symmetry breaking (SSB).

\section{Scalar potential and  spectrum} 
The most general $S_3$ invariant scalar potential involving three scalar
doublet fields is given by \cite{Chen:2004rr,Kubo:2004ps}
\begin{multline}
	\label{eqn:scalarpotential}
V = m^2\bigl(\phi_1^\dagger\phi_1+\phi_2^\dagger\phi_2\bigr)
+m_3^2\phi_3^\dagger\phi_3+\frac{\lambda_1}{2}\bigl(\phi_1^\dagger\phi_1
+\phi_2^\dagger\phi_2\bigr)^2
	+\frac{\lambda_2}{2}\bigl(\phi_1^\dagger\phi_1
-\phi_2^\dagger\phi_2\bigr)^2+\lambda_3\phi_1^\dagger\phi_2\phi_2^\dagger\phi_1+
\frac{\lambda_4}{2}\bigl(\phi_3^\dagger\phi_3)^2\\
+\lambda_5\bigl(\phi_3^\dagger\phi_3\bigr)
\bigl(\phi_1^\dagger\phi_1+\phi_2^\dagger\phi_2\bigr)
	+\lambda_6\phi_3^\dagger\bigl(\phi_1\phi_1^\dagger
+\phi_2\phi_2^\dagger\bigr)\phi_3
+\biggl[\lambda_7\phi_3^\dagger\phi_1\phi_3^\dagger\phi_2
+\lambda_8\phi_3^\dagger\bigl(\phi_1\phi_2^\dagger\phi_1
+\phi_2\phi_1^\dagger\phi_2\bigr)+ \text{h.~c.}\biggr] \, .
\end{multline}
After SSB, nine degrees of freedom are left: three neutral scalars,
two neutral pseudoscalars and two charged scalars with two degrees of
freedom each.  We denote the VEVs of $\phi_i$ by $v_i$ and assume the
$\lambda_i$'s to be real.  For the purpose of generating maximal
mixing in the lepton sector, we choose the vacuum alignment $v_1 = v_2
= v$.  Once we solve the tadpole equations, $v_1=v_2$ turns out to be
an extremal condition if the following relations are satisfied:
\begin{align}
-m^2 &= (2 \lambda_1+\lambda_3) v^2+ (\lambda_5+\lambda_6+\lambda_7)v_3^2
+3 \lambda_8 vv_3  \, , \nonumber\\
-m_3^2 &= \lambda_4 v_3^2+2 (\lambda_5+\lambda_6+\lambda_7) v^2
+2 \lambda_8 v^3/v_3 \, .
\end{align}
To ensure that the chosen vacuum alignment actually corresponds to a
minimum of the potential, we adjust parameters to make sure that the
determinant of the Hessian matrix is positive.  Just to recall, the
Hessian is defined as the square matrix of second order partial
derivatives of a function describing its local curvatures. In this
case, the function is the scalar potential and the Hessian is just the
mass matrix of the scalars.  The positivity of the eigenvalues -- see
later in Eq.~(\ref{eqn:scalarmasses}) -- guarantees that the potential
is minimized.  We note that after SSB, the potential turns out to be a
polynomial of order four, and its global stability in the asymptotic
limit (i.e. $\phi_i \to \infty$) is ensured by the following set of
conditions:
\begin{align}
\label{min_conditions} 
	\lambda_1 + \lambda_2 &> 0, & 
	\lambda_1 + \lambda_3 &> \lambda_2, & 
	\lambda_4 &> 0, & 
	\lambda_5 + \lambda_6 &> 0, & 
	\lambda_7 &> 0, &
	\lambda_8 &> 0.
\end{align}

We now set out to find the spectrum of the three CP-even neutral
scalars.  We insert the expansion $\phi_i^0=v_i+h_i$ in
Eq.~(\ref{eqn:scalarpotential}) to obtain the mass matrix. After its
diagonalization the weak basis scalars $h_{1,2,3}$ are expressed in
terms of the physical scalars $h_{a,b,c}$ as
\begin{align}
	\label{eqn:scalarmixing}
	h_1&= U_{1b}~h_b + U_{1c}~h_c - \frac{1}{\sqrt{2}} ~h_a \, , 
        \nonumber\\
	h_2&= U_{2b}~h_b + U_{2c}~h_c + \frac{1}{\sqrt{2}} ~h_a \, ,\\
	h_3&= U_{3b}~h_b + U_{3c}~h_c \, , \nonumber 
\end{align}
where $U_{ib}$ and $U_{ic}$ are analytically tractable but complicated
functions of $\lambda_i$s, $v$ and $v_3$, which we do not display.
The condition $v_1 = v_2$ immediately leads to $U_{1b}=U_{2b}$ and
$U_{1c}=U_{2c}$.  The masses of the three CP-even neutral scalars are
\begin{eqnarray}
	\label{eqn:scalarmasses}
  m_a^2 &=& 4\lambda_2 v^2 - 2\lambda_3 v^2 - v_3
\left(2\lambda_7 v_3 + 5\lambda_8 v\right)\, , \nonumber\\
  m_{b(c)}^2 &=& \frac{1}{2v_3}\left[4 \lambda_1 v^2 v_3+2
    \lambda_3 v^2 v_3+2 \lambda_4 v_3^3-2 \lambda_8 v^3+3
    \lambda_8 v v_3^2 \mp \Delta m^3\right]\, ;
\label{eqn:mabc}
\end{eqnarray}
where 
\begin{align}
\label{eqn:delta-m3}
\Delta m^3	=& \biggl[8 v v_3 \biggl\{2 v v_3^3 \biggl(2
   (\lambda_5+\lambda_6+\lambda_7)^2-\lambda_4 (2
   \lambda_1+\lambda_3)\biggr)+2 \lambda_8 v^4 (2 \lambda_1+\lambda_3)-3
   \lambda_4 \lambda_8 v_3^4\nonumber\\
&+12 \lambda_8 v^2 v_3^2
   (\lambda_5+\lambda_6+\lambda_7)+12 \lambda_8^2 v^3
   v_3\biggl\}
+\biggl\{2 v^2 v_3 (2 \lambda_1+\lambda_3)+2
   \lambda_4 v_3^3-2 \lambda_8 v^3+3 \lambda_8 v
   v_3^2\biggr\}^2\biggr]^\frac{1}{2}\, .
\end{align}

\begin{figure}[htb]
	\begin{center}
		\includegraphics[width=5.5cm]{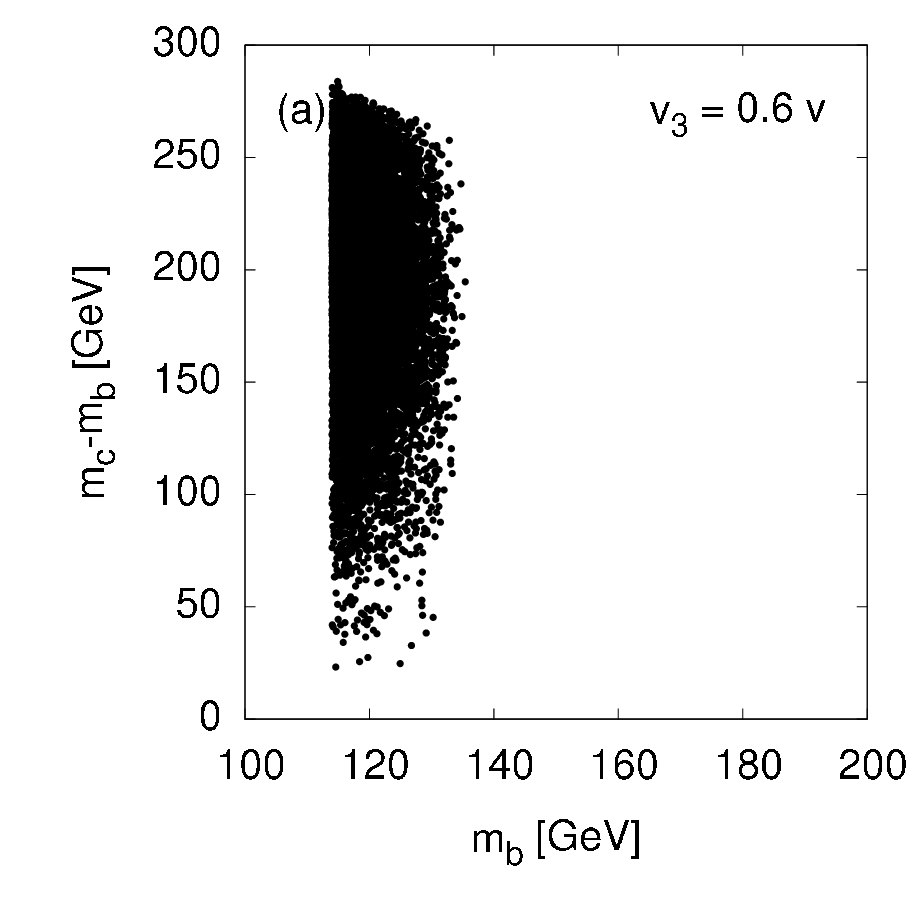}
			\includegraphics[width=5.5cm]{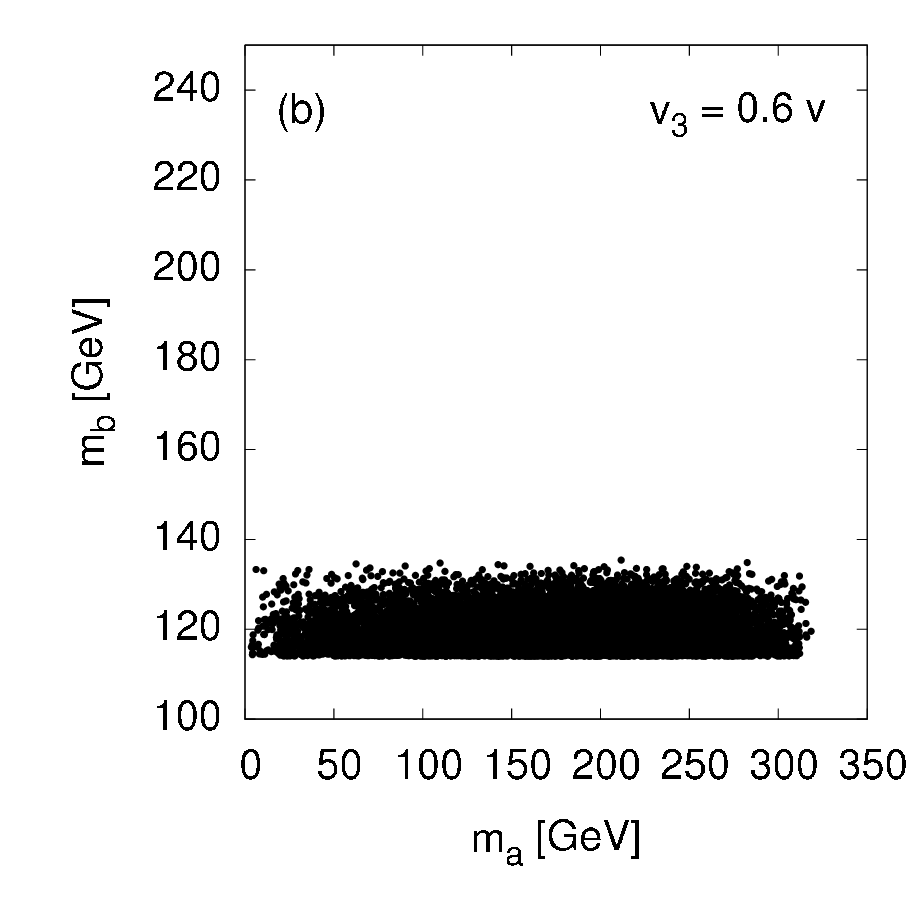}
                        \caption{\small{\sf Results of a random search for
                            allowed scalar masses for a fixed $v_3/v=0.6$. In
                            the left panel (a), we exhibit the splitting
                            $(m_c-m_b)$ for different choices of $m_b$. In the
                            right panel (b), we show the allowed range of
                            $m_a$.}}
	\label{fig:scalar_massrelations}
\end{center}
\end{figure}

A few things are worth noting at this stage:
\begin{enumerate} 
\item[$(i)$] Since $\phi_{1,2,3}$ are all weak $SU(2)$ doublets, their VEVs
  are related as: $2 v^2 + v_3^2 = v_{\text{SM}}^2$, where $v_{\text{SM}} \approx 246~
  \text{GeV}$.

\item[$(ii)$] One of the physical scalars is given by $h_a=(h_2-h_1)/\sqrt{2}$,
  i.e. there is no dependence on $\lambda_{\{1,\ldots,8\}}$ or on the
  VEVs. This happens because $S_3$ symmetry requires the scalar mass matrix to
  be of the form
\begin{align*}
	\begin{pmatrix}
		a&	b&	c\\
		b&	a&	c\\
		c&	c&	d\\
	\end{pmatrix}\, ,
\end{align*}
which always yields $(-1,1,0)$ as one eigenvector, regardless of the
values of $a,b,c$ and $d$.

\item[$(iii)$] We strictly follow Eq.~(\ref{min_conditions}) to ensure
  that the potential is bounded from below. We randomly vary the {\em
    magnitude} of the $\lambda_i$'s in the range $[0,1]$, although
  slightly larger (but $<4\pi$) values of $|\lambda_i|$ would have
  still kept the couplings perturbative.  We {\em accept} a given set
  of $\{\lambda_1, \ldots, \lambda_8, v\}$ only if it satisfies the
  minimization conditions.

\item[$(iv)$] The difference $m_c^2 - m_b^2 = \Delta m^3/v_3$ is
  positive, and when $v_3 \to 0$, i.e. $v \to v_\text{SM}$, the
  splitting grows enormously. Since the maximum value of $m_c^2$ is
  controlled by $\lambda_i \leq 1$, $m_b^2$ becomes tachyonic when
  $v_3 \to 0$.  It has been suggested in \cite{Chen:2004rr} that with
  {\em order one} Yukawa couplings, the ratio $v_3/v \sim 0.1$
  reproduces the correct Cabibbo angle in the quark sector. We require
  $v_3/v \geq 0.6$ to ensure that $m_b^2$ stays above the accepted
  limit.  Since $h_b$ and $h_c$ have similar gauge and Yukawa
  properties, quite different from those of $h_a$ (see discussions
  later), we show the mass splitting $(m_c - m_b)$ against $m_b$ in
  Fig.~\ref{fig:scalar_massrelations}(a), and the relation between
  $m_b$ and $m_a$ in Fig.~\ref{fig:scalar_massrelations}(b).

\end{enumerate}

\section{Scalar couplings to gauge and matter fields}
The kinetic terms $\left|D_\mu \phi_i\right|^2$ ($i=1,2,3$) yield the
couplings of the symmetry basis $h_i$ to $W^\pm$ and $Z$. Clearly,
these couplings are modified by a factor of $v_{i}/v_\text{SM} < 1$
compared to their SM expressions.  In terms of the mass basis scalars,
we observe the following: ($i$) The coupling of $h_b$ to $W^+W^-$ (or,
$ZZ$) is the corresponding SM coupling multiplied by $\left(2v U_{1b}
  + v_3 U_{3b}\right)/v_\text{SM}$ and the corresponding factor for
$h_c$ is $\left(2v U_{1c} + v_3 U_{3c}\right)/v_\text{SM}$.  ($ii$)
The scalar $h_a$ does not have $h_aZZ$ or $h_aWW$ couplings, unlike
the other two scalars. This can be understood as follows. The gauge
couplings of $h_i$ arise from the linear expansion $\phi_i^0 = v_i +
h_i$ in the kinetic term. Since $v_1 = v_2 = v$, the combination
$(h_1+h_2)$ will couple to gauge bosons as proportional to $v$. The
orthogonal combination $(h_2 - h_1)$ that represents the physical
scalar $h_a$ -- see Eq.~(\ref{eqn:scalarmixing}) and point $ii$
following Eq.~(\ref{eqn:delta-m3}) -- will not have the usual
scalar-gauge-gauge vertex. The four-point $h_a^2ZZ$ and $h_a^2WW$
couplings will, however, exist.

The $S_3$ invariant Yukawa Lagrangian, where the neutral scalars and
the charged leptons/quarks are in their weak basis, is given by
($f_i$'s are the leptonic and $g_i^{u/d}$s are quark Yukawa couplings)
\begin{eqnarray}
	\mathcal{L}_{\text{Yuk}} &=& f_4 e e^c h_3 + f_5 e \mu^c h_3 
+ f_1 \mu^c (\mu h_2 + \tau h_1) + f_2 \tau^c (- \mu h_2 + \tau h_1)
\nonumber \\
&\quad& +g_4^u u u^c h_3 + g_5^u u c^c h_3 
+ g_1^u c^c (c h_2 + t h_1) + g_2^u t^c (- c h_2 + t h_1) \\
&\quad&  + g_4^d d d^c h_3 + g_5^d d s^c h_3 
+ g_1^d s^c (s h_2 + b h_1) + g_2^d b^c (- s h_2 + b h_1)
+ \text{H.~c.} \nonumber 
\end{eqnarray}
The couplings of $h_{b,c}$ to the quarks and leptons depend on the parameters
$v,v_3, \lambda_i$ and $f_i$ (or, $g_i^{u/d}$), while the couplings of $h_a$
to fermions depend only on $f_i$ (or, $g_i^{u/d}$).  The {\em physical}
scalar couplings to the {\em mass basis} fermions are given by the following Yukawa
matrices, displayed for the charged leptons as an example (the structures for
the quark sector are similar {\em modulo} Cabibbo mixing):
\begin{align}
	\label{eqn:yukawas}
	Y_{h_a} &= 
	\begin{pmatrix}
 0        &	0     &Y_{e_L\tau_R}^a \\
 0        &	0    &Y_{\mu_L\tau_R}^a \\
 Y_{\tau_L e_R}^a &	Y_{\tau_L\mu_R}^a&0        \\ 
	\end{pmatrix}, & 
	Y_{h_{b,c}} &= 
	\begin{pmatrix}
 Y_{e_Le_R}^{b,c}        &	Y_{e_L\mu_R}^{b,c}     &0 \\
 Y_{\mu_L e_R}^{b,c}        &	Y_{\mu_L\mu_R}^{b,c}    &0 \\
 0 &	0&Y_{\tau_L\tau_R}^{b,c}        \\ 
	\end{pmatrix} \, .
\end{align}
The position of the zeros in the matrices deserves some attention.  It turns
out that $h_{a,b,c}$ have off-diagonal fermion couplings at tree level due to
the absence of any natural flavor conservation \cite{Glashow:1976nt}. The
numerical entries of the Yukawa matrices are intimately tied to the successful
reproduction of the quarks' and leptons' masses and mixings.  We make three 
observations at this stage.
$(i)$~ We admit that unless some $S_3$ breaking parameters are
  introduced a successful reproduction of $V_{cb}$ and $V_{ts}$ is
  problematic \cite{Xing:2010iu}, and also domain walls will be
  formed. We do not intend to cover all flavor issues. We simply
  concentrate on the scalar sector whose Lagrangian is $S_3$ invariant
  to start with.
$(ii)$~ $h_a$ couples {\em only} off-diagonally and one of the two
  fermions has to be from the third generation.
$(iii)$~ $h_{b,c}$ couple diagonally as in the SM, {\em but also} possess
  small, numerically insignificant, off-diagonal couplings involving the first
  two generations.

The last two points require further clarification. In a theory with
more than one $SU(2)$ Higgs doublet, tree level flavor changing
neutral currents (FCNC) generally exist in the scalar sector. For
example, they exist in the ordinary `two Higgs doublet model (2HDM)',
but in the supersymmetric standard model they are avoided by the
arrangement that one doublet couples only to the up-type fermions and
the other to only the down-types.  In nonsupersymmetric scenarios, in
the absence of any natural flavor conservation, symmetry arguments
have been advanced in the context of multi-Higgs models to show that
the off-diagonal Yukawa couplings of the neutral scalars are
suppressed by their relation to the off-diagonal entries of the
Cabibbo-Kobayashi-Maskawa (CKM) matrix \cite{Botella:2009pq}.

In the present case, $S_3$ symmetry, under which both scalars and
fermions transform nontrivially, is instrumental in suppressing the
off-diagonal couplings. To provide intuitive understanding, we take,
as an example, only the two-flavor $\mu$--$\tau$ sector together with
two neutral scalars $h_1$ and $h_2$. It is not difficult to see that
the combination $(h_2-h_1)$, which corresponds to $h_a$, couples only
off-diagonally, as mentioned earlier. But the other combination
$(h_2+h_1)$, which corresponds to $h_{b,c}$ following
Eq.~(\ref{eqn:scalarmixing}), couples only diagonally to physical
$\mu$ or $\tau$. When we consider the quark sector, $\mu$ and $\tau$
would be replaced by second and third generation quarks which will
have CKM mixing.  This will yield off-diagonal entries for $h_{b,c}$
couplings to quarks suppressed by the off-diagonal CKM elements.  The
same happens for off-diagonal couplings involving the first two
generations as well.  The tiny size of tree level FCNC rates in an
$S_3$ flavor model has been noticed also earlier, where predictions
for ${\rm Br}~(\tau \to 3\mu)$, ${\rm Br}~(K_L \to 2 e)$ and ${\rm
  Br}~(B_s \to 2 \mu)$ have been given \cite{Kubo:2003iw}.  In some
setups where the fermion transformations under $S_3$ are not
appropriately adjusted, the off-diagonal Yukawa couplings may become
order one which induce sizable neutral scalar mediated rare processes,
like $K_L \to \mu e$ or $K_L \to 2\pi$, at tree level.  This requires
those neutral scalars to lie beyond several TeV
\cite{Yamanaka:1981pa,Kubo:2004ps}. But in our case, once we adjust
the $f_i/g_i^{u,d}$'s to reproduce the fermion masses and mixing, the
off-diagonal Yukawa couplings are determined too.  The largest of them
corresponds to $\bar{c}_L t_R h_a$, which is about 0.8. The second
largest off-diagonal coupling is that for $\bar{s}_L b_R h_a$, and is
about 0.02. The next in line is $\bar{\mu}_L \tau_R h_a$, whose
coefficient is about 0.008.  The others are orders of magnitude
smaller, and are of no numerical significance. Although FCNC processes
like $B_d$--$\bar{B}_d$ and $B_s$--$\bar{B}_s$ mixings proceed at tree
level, the contributions are adequately suppressed even for light
scalar mediators.

\section{Collider signatures}
The perturbativity condition $\left|\lambda_i\right| \leq 1$ and the
requirement $m_{b/c} \geq 114$\,GeV (for which we set $v_3/v \simeq
0.6$) yields $m_b$ in the neighbourhood of 120\,GeV and $m_c$ within
400\,GeV -- see the scatter plots in
Fig.~\ref{fig:scalar_massrelations}.  Both $h_b$ and $h_c$ would decay
into the {\em usual} $ZZ$, $WW$, $b\bar b$, $\gamma\gamma$, $\cdots$
modes, but the dominant decay mode of $h_b$ (or $h_c$) for the case of
$m_a< m_b/2$ (or $m_a < m_c/2$) would be into $h_a h_a$. Recall that
the existing limits on the Higgs mass depend crucially on the gauge
coupling of the Higgs.  Since $h_aZZ$ or $h_aWW$ couplings are
nonexistent, the mass of $h_a$ is unconstrained, i.e. $m_a$ can be
lower than 114\,GeV or larger than 200\,GeV.  We numerically calculate
the strength of the $h_bh_ah_a$ coupling from the set of {\em
  acceptable} parameters characterizing the potential, and introduce a
parameter $k$ which is the ratio of the $h_bh_ah_a$ coupling and the
$h_bWW$ coupling. The magnitude of $k$ depends on the choice of
$\lambda_i$ and $v_3$. Assuming $m_a = 50$\,GeV, we obtain $k$ in the
range of $(5-30)$. Just to compare with a 2HDM \cite{LopezVal:2009qy}
for illustration, the corresponding $k$ value, when the heavier Higgs
weighing around 400\,GeV decays into two lighter Higgs weighing
114\,GeV each, is about 10.

\begin{figure}[tb]
\begin{center}
\includegraphics[width=5.5cm]{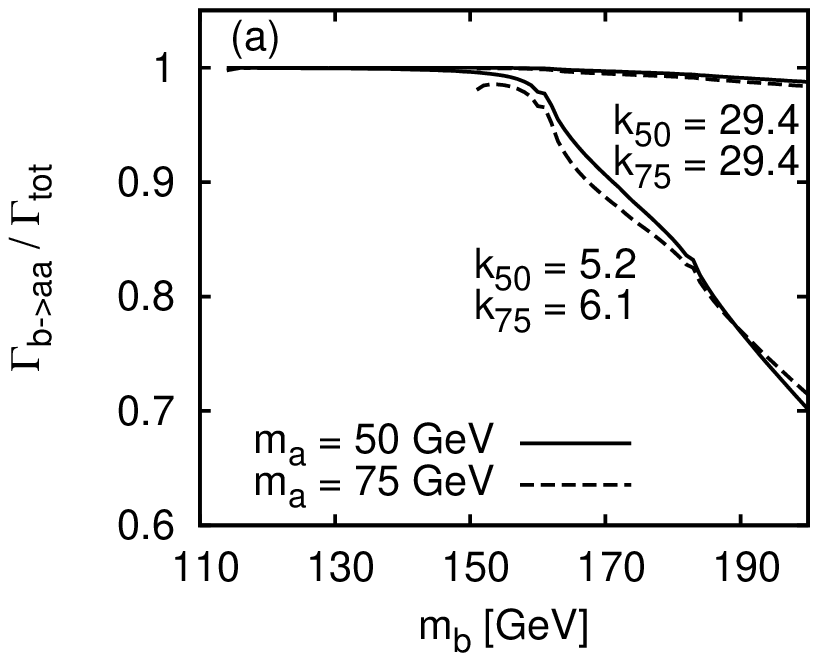}
\includegraphics[width=5.5cm]{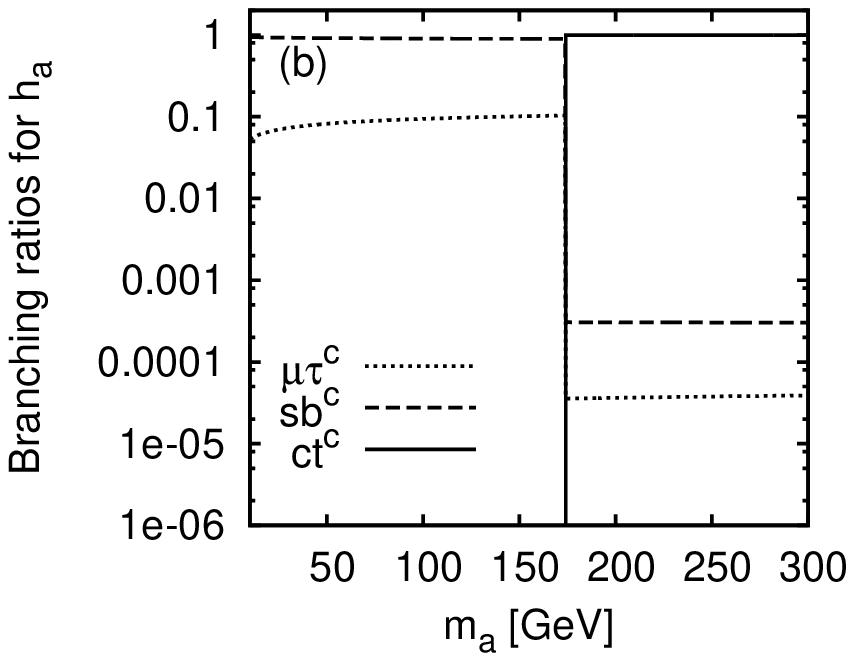}
\includegraphics[width=5.5cm]{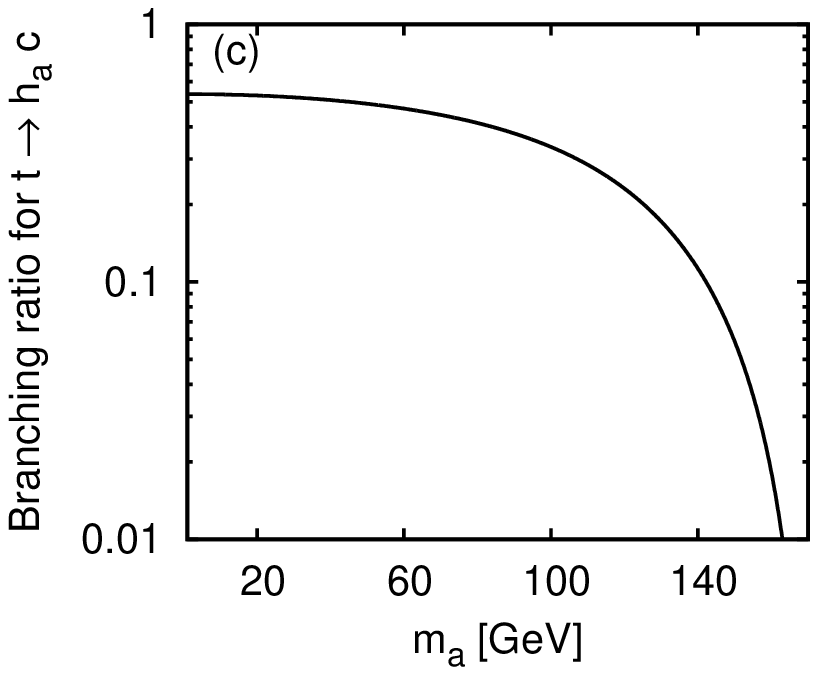}
\caption{\small{\sf (a) Branching ratio of $h_b\to h_ah_a$ for two
    representative values of $m_a$, and in each case for smallest / largest
    values of $k$ in the set of accepted scalar parameters which compares the
    strength of the $h_bh_ah_a$ coupling to the strength of $h_bWW$ coupling;
    (b) branching ratios for the decay of $h_a$, and (c) branching ratio of
    the top quark decay into $h_a$ and charm quark.}}
\label{fig:decays}
\end{center}
\end{figure}

In Fig.~{\ref{fig:decays}}(a) we have plotted the branching ratio of
$h_b \to h_a h_a$ as a function of $m_b$ for two representative values
$m_a = 50$,$75$\,GeV, and for $k \sim$ 5 and 30, which correspond to
the smallest and largest $k$ obtained from the set of
\textit{accepted} scalar parameters.  We observe that till the $WW$ or
$ZZ$ decay modes open up, the branching ratio $h_b \to h_ah_a$ is
almost 100\%.  To calculate the decay widths into the usual modes
(other than $h_ah_a$), we have used \texttt{HDECAY}
\cite{Djouadi:1997yw} by appropriately modifying the gauge and Yukawa
couplings.  

As Fig.~\ref{fig:decays}(b) suggests, as long as $m_a < m_t$, $h_a$
will dominantly decay into jets, and one of them can be identified as
the $b$-jet.  The branching ratio of $h_a \to \mu\bar\tau$ is,
nevertheless, not negligible (about 0.1). As shown in
Fig.~\ref{fig:decays}(c), for $m_a \ll m_t$, the branching ratio of $t
\to h_a c$ is quite sizable, which falls with increasing $m_a$. It may
be possible to reconstruct $h_a$ from $h_a \to \mu \bar\tau$.  In
fact, a light $h_a$ would be copiously produced from the top decay at
the LHC.  On the other hand, if $m_a > m_t$, as can be seen again from
Fig.~\ref{fig:decays}(b), $h_a$ decays to $t\bar c$ with an almost
100\% branching ratio.

If $k$ is large, then there is an interesting twist to the failed
Higgs search at LEP-2. In this case, $h_b \to h_a h_a$ would overwhelm
$h_b \to b\bar b$, and hence the conventional search for the SM-like
scalar ($h_b$, as the lighter between $h_b$ and $h_c$) would fail.
This is similar to what happens in the next-to-minimal supersymmetric
models, when the lightest scalar would dominantly decay into two
pseudoscalars, and each pseudoscalar would then decay into $2b$ or
$2\tau$ final states. In view of these possible $4b$ or $4\tau$ Higgs
signals, LEP data have been reanalyzed putting constraints on the
Higgs production cross section times the decay branching ratios
\cite{Schael:2006cr,Schael:2010aw}.  The possibility of the Higgs
cascade decays into $4j$ ($j=$ quark/gluon), $2j+2$ photons and 4
photons has been studied too \cite{Chang:2005ht,Chang:2006bw}. From a
study of $4b$ final states, a limit $m_h > 110$ GeV (for a SM-like
Higgs) has been obtained \cite{Chang:2005ht}.  From all other cascade
decays the limit on $m_h$ will be considerably weaker.  Our $h_a$ has
the special feature that it has only off-diagonal Yukawa couplings
involving one third-family fermion. If $h_b$ is lighter than the top
quark, it would decay as $h_b \to h_a h_a \to 2b+2j$, and into
$b+1j+\mu+\tau$, the latter constituting a spectacular signal with two
different lepton flavors $\mu$ and $\tau$. The standard $2b$ and
cascade $4b$ decay searches are not sensitive to our final states, and
so a value of $m_b$ much lighter than 110 GeV is not ruled out.

\section{Conclusions}
The discrete flavor symmetry $S_3$, besides successfully reproducing
fermion masses and mixing, provides an extended Higgs sector having
unconventional decay properties. We assume all the couplings to be
real, and do not deal with the possibility of CP violation in this
paper.  The potential has been minimized requiring maximal mixing for
the atmospheric neutrinos. In our setup, there are two scalars which
are SM Higgs like, {\em except} that each of them can have a dominant
decay into the third ($h_{b,c} \to h_a h_a$). The latter, i.e. $h_a$,
has no $h_a VV$-type gauge interactions, and has {\em only} flavor
off-diagonal Yukawa couplings with one fermion from the third
generation. It is not unlikely that by evading the conventional search
strategies, both $h_b$ and $h_a$ are already buried in the existing
LEP and Tevatron data.  In this analysis we have not dealt with the
two pseudoscalars, which we leave for a future study.  We urge our
experimental colleagues to look for our suggested signals at the LHC,
and perhaps also reanalyze the existing collider data.

\noindent {\bf{Acknowledgments:}}~ We thank M. Frigerio for useful
discussions. This work was supported by DAAD-DST PPP Grant
No.~D/08/04933, and DST-DAAD project
No.~INT/DAAD/P-181/2008. G.B. acknowledges hospitality at
T.U. Dortmund, IST-Lisbon, LPT-Orsay, and ICTP-Trieste, at different
stages of this work. H.P. was supported by DFG Grant No. PA
803/6-1. P.L. and H.P. were also supported by the Physics at the
Terascale Helmholtz Alliance Working Group: Neutrino masses and Lepton
Flavor Violation at the LHC, and they acknowledge hospitality at the
Saha Institute of Nuclear Physics, Kolkata, during a part of this
collaboration.

\end{document}